\documentclass[conference]{IEEEtran}

\usepackage{cite}
\usepackage[cmex10]{amsmath}
\usepackage{tabularx,colortbl}
\usepackage{amsmath,amssymb,amsthm,mathrsfs,dsfont}

\usepackage{amsmath}
\usepackage{amssymb}

\usepackage{booktabs}
\usepackage{threeparttable}
\usepackage{multirow}

\usepackage{algorithm}
\usepackage{algpseudocode}

\usepackage{subfigure}
\usepackage{stfloats}

\newcounter{MYtempeqncnt}
\IEEEoverridecommandlockouts
\ifCLASSINFOpdf
  \usepackage[pdftex]{graphicx}
  \DeclareGraphicsExtensions{.pdf,.jpeg,.png}
\else
  \usepackage[dvips]{graphicx}
  \DeclareGraphicsExtensions{.eps}
\fi

\begin{document}

\title{\huge{Accumulate then Forward: An Opportunistic Relaying Protocol for Wireless-Powered Cooperative Communications}}

\author{\IEEEauthorblockN{Ziyi Li, He (Henry) Chen, Yifan Gu, Yonghui Li, and Branka Vucetic}
\IEEEauthorblockA{School of Electrical and Information Engineering,
University of Sydney, Sydney, NSW 2006, Australia\\
E-mail:~\{ziyi.li,~he.chen,~yifan.gu,~yonghui.li,~branka.vucetic\}@sydney.edu.au
}
}

\maketitle
\begin{abstract}
This paper investigates a wireless-powered cooperative communication network consisting of a source, a destination and a multi-antenna decode-and-forward relay. We consider the relay as a wireless-powered node that has no external power supply; but it is equipped with an energy harvesting (EH) unit and a rechargeable battery such that it can harvest and accumulate energy from radio-frequency signals broadcast by the source. By fully incorporating the EH feature of the relay, we develop an opportunistic relaying protocol, termed accumulate-then-forward (ATF), for the considered WPCCN. 
We then adopt the discrete Markov chain to model the dynamic charging and discharging behaviors of the relay battery. Based on this, we derive a closed-form expression for the exact outage probability of the proposed ATF protocol. Numerical results show that the ATF scheme can outperform the direct transmission one, especially when the amount of energy consumed by relay for information forwarding is optimized.


\end{abstract}
\IEEEpeerreviewmaketitle

\section{Introduction}
%
%
Radio-frequency (RF) energy harvesting technique has recently been regarded as a new viable solution to extend the lifetime of energy-constrained wireless networks \cite{xiao2014wireless}. This technique
has opened up new opportunities for cooperative communications as it enables a new cooperative manner for wireless devices (see \cite{bi2015wireless} and references therein). In particular, the relay in conventional cooperative networks, can now harvest energy from the information source and then use the harvested energy to assist the source's transmission. In this sense, the relay is more willing to cooperate since it does not need to consume its own energy. In this paper, we refer to a cooperative communication network with wireless powered relay as a wireless-powered cooperative communication network (WPCCN). In fact, the design and analysis of WPCCNs have become a hot research topic in wireless communication area very recently (see, e.g., \cite{krikidis2012rf,chen2015harvest,krikidis2015relay,liu2015performance,gu2016distributed,nasir2015block}). It is worth pointing out that the design of wireless-powered networks is inherently different from that of conventional wireless networks powered by solar/wind energy harvesting (e.g., \cite{michelusi2013transmission}). Specifically, in wireless-powered networks, the amount of energy harvested by wireless-powered nodes highly depends on the network operation modes, while these two events are independent in conventional energy harvesting networks.


In practice, due to the propagation loss of RF signals, the amount of energy harvested by wireless-powered\footnote{Throughout this paper, we use the terms ``wireless-powered" and ``energy harvesting" interchangeably.} nodes during one transmission block are normally very limited. Thus, it is necessary for these nodes to equip with energy storage (e.g., a rechargeable battery) such that they can accumulate enough amount of harvested energy before performing one round of information transmission. However, to the best knowledge of the authors, only a few papers in open literature incorporated the energy accumulation process in the design/analysis of WPCCNs. In \cite{krikidis2012rf}, Krikidis \emph{et al.} studied a classical three-node relay network with an energy harvesting relay, where the relay was assumed to have a discrete and finite-capacity rechargeable battery and the energy accumulation process at the relay was characterized by a finite-state Markov chain. Recently, \cite{krikidis2015relay,liu2015performance,gu2016distributed} extended \cite{krikidis2012rf} to a more general scenario with multiple energy harvesting relays, in which several relay selection schemes were proposed and analyzed. Besides, a continuous battery model was adopted in \cite{nasir2015block} to analyze the throughput performance of a three-node relay network with energy accumulation. However, an infinite capacity of  the relay battery was assumed in \cite{nasir2015block}, which makes the resulting analysis somewhat ideal.


A common assumption in \cite{krikidis2012rf,krikidis2015relay,liu2015performance,gu2016distributed,nasir2015block} is that no direct link exists between source and destination. However, the direct link of a WPCCN actually plays a crucial role in practice. This is because the amount of energy harvested by the wireless-powered relay is generally limited, which means the signal-to-noise ratio (SNR) of the source-destination link may be comparable or even larger than that of the relay-destination link. In other word, relying solely on an energy harvesting relay to accomplish the information delivery from source to destination may lead to poor system performance. Thus, the direct link should be incorporated in the designs of WPCCNs. When the direct link is available, a natural question that arises is ``\emph{how often should the source cooperate with the energy harvesting relay}?". This is actually a non-trivial problem. Specifically, the relay can only accumulate little amount of energy for information forwarding when the cooperation is too intensive, which may lead to even worse performance than the direct transmission (i.e., no cooperation) scheme. On the other side, few cooperation will yield the under-utilization of the relay and then make the cooperation gain insignificant.

Motivated by this open problem, in this paper we focus on the design of a three-node WPCCN consisting of one source, one energy harvesting relay implementing decode-and-forward protocol and one destination, where the direct link between source and destination exists. We consider that the relay is equipped with a rechargeable battery and multiple antennas, which includes single-antenna relay scenarios studied in \cite{krikidis2012rf,krikidis2015relay,liu2015performance,nasir2015block} as special cases. Towards the appropriate usage of the energy harvesting relay, we develop an opportunistic relaying protocol, termed accumulate-then-forward (ATF), for the considered WPCCN, where the relay cooperates with the source in an opportunistic manner. Particularly, the cooperation between source and relay is activated only when the accumulated energy at relay exceeds a predefined energy threshold and the decoding of source's information at relay is successful.
Otherwise, the source has to transmit information to destination by itself, while the relay accumulates the energy harvested from source's signals.
By modeling dynamic charging/discharging behaviors of relay battery as a finite-state Markov chain, we then analyze the outage probability of the proposed ATF protocol over mixed Rician-Rayleigh fading channels. Specifically, in contrast to the Rayleigh fading model used in \cite{krikidis2012rf,krikidis2015relay,liu2015performance,nasir2015block}, we adopt Rician fading to characterize the channel fading between source and relay. This is more practical by considering that the line-of-sight (LoS) path is very likely to exist between source and relay as the current wireless energy harvesting techniques can only support a relatively short distance. Numerical results are finally performed to verify the theoretical analysis and illustrate the effects of several system parameters on the network outage probability.

%

\section{System Model and Protocol Design}\label{System}
We consider a WPCCN consisting of one single-antenna source $S$, one decode-and-forward (DF) relay $R$ equipped with $N$ antennas, and one single-antenna destination $D$. We assume that $S$ and $D$ have embedded power supplies, while $R$ is a wireless-powered device. But, $R$ is equipped with an wireless energy harvesting unit and a rechargeable battery such that it can accumulate the energy harvested from RF signals broadcast by $S$. Furthermore, the relay is equipped with separate energy and information receivers \cite{zhang2013mimo}. As such, it can flexibly switch its received signal to one of these two receivers to realize energy harvesting or information~decoding.

Let $h_{SD}$ denote the complex channel coefficient of $S$-$D$ link. Also let $\mathbf{h}_{SR}$ and $\mathbf{h}_{RD}$ denote the $N\times1$ channel vectors of $S$-$R$ and $R$-$D$ link, respectively. As the up-to-date wireless energy harvesting techniques could only be operated within a relatively short distance, the line-of-sight (LoS) path is very likely to exist between $S$ and $R$. Motivated by this fact, we consider an asymmetric scenario for the fading distributions of $S$-$R$ link and $R$-$D$ link. In particular, the elements of $\mathbf{h}_{SR}$ are subject to independent and identically distributed (i.i.d.) Rician fading, while the elements of $\mathbf{h}_{RD}$ are subject to i.i.d. Rayleigh fading. Besides, the channel coefficient of direct link $h_{SD}$ follows Rayleigh distribution. Furthermore, all channels between $S$, $R$ and $D$ are assumed to experience slow, independent and frequency flat fading such that the channel gains remain unchanged within each transmission block but change independently from one block to the other.

We now propose an accumulate-then-forward (ATF) protocol for WPCCN. In the proposed ATF protocol, the energy harvesting relay accumulates the energy harvested from signals broadcast by $S$ and assists its information transmission in an opportunistic manner. Specifically, the relay opportunistically switches between the energy harvesting mode and the information cooperation mode based on both whether its residual energy in battery exceeds a predefined energy threshold and whether its decoding of source's information is correct.
Here, we use $T$ to denote the duration of each transmission block, which is further divided into two time slots with equal length $T/2$ when $R$ attempts to decode source's information (i.e., the residual energy at $R$ exceeds the predefined energy threshold).
\begin{itemize}
  \item \textbf{Mode I}: This mode corresponds to the case when the current residual energy in relay battery, denoted by $E_R$, is less than the energy threshold $E_T$. In this situation, $R$ chooses to perform energy harvesting operation to further accumulate its energy and the source has to transmit its information to $D$ by itself during the whole transmission~block.
  \item \textbf{Mode II}: In this mode, the current residual energy $E_R$ at $R$ is not less than $E_T$, but the decoding of the source's information at $R$ during the first time slot is unsuccessful. Thus, $R$ still cannot cooperate with $S$ in this mode. During the second time slot, $S$ is motivated to re-transmit its information to enhance the received SNR at $D$, while $R$ can harvest energy from this signal to charge its battery.
  \item \textbf{Mode III}: Here, $R$ has accumulated enough energy (i.e., $E_R \ge E_T$) and its information decoding of the first hop is also correct. As such, $R$ will work in information cooperation mode by helping forward the source's information to $D$ during the second time slot, while $S$ can keep in silence during this period.
\end{itemize}

In the following, we express the harvested energy at $R$ and the received SNR at $D$ of the proposed ATF for three possible modes mathematically. Without lose of generality, we consider a normalized transmission block (i.e., $T=1$) hereafter. Moreover, we use $P_S$ to denote the transmit power of $S$ and $H_{xy}=\|\mathbf{h}_{xy}\|^2$ to denote the channel power gain, where $x,y\in\left\{S,R,D\right\}$ and $\|\mathbf{x}\|$ denotes the Euclidean norm of a vector $\mathbf{x}$.

\subsection{Mode I}
In this mode, $R$ harvests energy during the whole transmission block. Thus, the amount of harvested energy at $R$ can be expressed as
\begin{equation}\label{eq:harvest_energy_I}
E_{H}^{\rm{I}} = \eta P_S H_{SR},
\end{equation}
where $\eta\in\left(0,1\right]$ is the energy conversion efficiency. In (\ref{eq:harvest_energy_I}), we ignore the amount of energy harvested from the noise as it is normally below the sensitivity of energy harvesting circuit.

Let $x_S$ denote the transmitted signal by $S$ with unit energy. In Mode I, the received signal at $D$ comes only from the direct link.
Thus, the received SNR at $D$ in Mode I is given~by
\begin{equation}
\gamma_{D}^{\rm{I}} = \gamma_{SD}= P_S H_{SD}/N_0,
\end{equation}
where $N_0$ is the power of the additive Gaussian white noise (AWGN).

\subsection{Mode II}
Since $R$ has stored sufficient energy in Mode II, during the first time slot, it will try to perform information decoding based on the signal received from $S$.
 We assume that the maximum ratio combining (MRC) technique is adopted at $R$ to maximize the received SNR. In this case, the received SNR at the $R$ is given~by
\begin{equation}
\gamma_{SR} = P_S H_{SR}/N_0.
\end{equation}

Recall that in this mode the source's information is not decoded correctly at $R$. In the second time slot, $S$ has to re-transmit its information to $D$ as $R$ cannot help, while $R$ can harvest energy from this re-transmitted signal. The harvested energy during the second time slot is given by
\begin{equation}
E_{H}^{\rm{II}} = \eta P_S H_{SR}/2.
\end{equation}
On the other hand, $D$ receives two copies of the same information from $S$. With the MRC technique, the received SNR at $D$ in Mode II can be expressed as
\begin{equation}
\gamma_{D}^{\rm{II}} = \gamma_{SD} + \gamma_{SD} = 2 P_S H_{SD}/N_0.
\end{equation}

\subsection{Mode III}
In this mode, $R$ utilizes the received signal from $S$ to decode the information in the first hop and the decoding is correct. Thus, $R$ can cooperate with $S$ by forwarding the source's information to $D$. Here, we consider that $R$ will spend $E_T$ amount of energy to perform information forwarding and the fixed transmit power is assumed to be $P_R = 2 E_T$. Moreover, for simplicity, it employs transmit antenna selection scheme \cite{tannious2008spectrally}. That is, the antenna with maximum channel power gain is selected to forward information. We define $h_{RD}^* = \max \{\mathbf{h}_{RD}\}$. In this case, the received SNR at $D$ during the second time slot is given by
\begin{equation}
\gamma_{RD} = P_R H_{RD}/N_0,
\end{equation}
where $H_{RD}=|h_{RD}^*|^2$ is the channel power gain of $R$-$D$ link. Note that the received SNR at $D$ during the first time slot is same as (2). Using the MRC technique, the resulting SNR at $D$ in Mode III can be characterized as
\begin{equation}
\gamma_{D}^{\rm{III}} = \gamma_{SD} + \gamma_{RD} = \left(P_S H_{SD} + P_R H_{RD}\right)/{N_0}.
\end{equation}

\section{Outage Probability Analysis}\label{Analysis}
In this section, we analyze the outage probability of the proposed ATF protocol over mixed Rician-Rayleigh fading channels. To this end, we first model the dynamic behaviors of the relay battery by a finite-state Markov chain (MC)~\cite{krikidis2012rf}.

\subsection{Markov Chain Description}
We assume that $R$ is equipped by a $L$ discrete-level battery with a capacity $C$. The $i$th energy level is defined as $\varepsilon_i = i C / L$, $i\in \{0,1,2,\ldots,L \}$. We define the state $S_i$ as the state of relay's residual battery being $\varepsilon_i$. $P_{i,j}$ is defined as the state transition probability from $S_i$ to $S_j$. Let $\lambda[m]\in\{\lambda_{{\rm{I}}},\lambda_{{\rm{II}}},\lambda_{{\rm{III}}}\}$ denote the system operation mode in the $m$-th transmission block, where $\lambda_X$, $X\in{\left({\rm{I}},{\rm{II}},{\rm{III}}\right)}$, represents the event that the $X$-th mode is operated. Considering the discrete battery model adopted in this paper, the discretized amount of energy $\varepsilon_H$ harvested by $R$ should be re-calculated as
\begin{equation}\label{varepsilon_H}
\small{
\varepsilon_H^X \triangleq \varepsilon_i,~{\rm{where}}~ i = \arg\max\nolimits_{j\in\{0,1,\ldots,L\}}\left\{\varepsilon_j:\varepsilon_j<E_H^X\right\},
}\end{equation}
where $X\in\{{\rm{I}},{\rm{II}}\}$. Similarly, the actual amount of energy consumed by $R$ for information forwarding should be defined~by
\begin{equation}\label{varepsilon_T}
\varepsilon_T \triangleq \varepsilon_i,~{\rm{where}}~ i = \arg\min\nolimits_{j\in\{1,\ldots,L\}}\left\{\varepsilon_j:\varepsilon_j\geq E_T\right\}.
\end{equation}
In this paper, we assume that $R$ can decode the information correctly if its received SNR exceeds a predetermined threshold. Let $\mathbb{R}$ denote the transmit rate of $S$. The SNR threshold of $S$-$R$ link in Mode II or III can then be defined as $\gamma_0 = 2^{2\mathbb{R}}-1$. We now can describe the three possible operations of the proposed ATF protocol during the $m$-th transmission block mathematically as follows
\begin{equation}
\lambda[m] =
\begin{cases}
\lambda_{\rm{I}}, &\mbox{if ~$\varepsilon_T>\varepsilon[m]$},\\
\lambda_{\rm{II}}, &\mbox{if ~$\varepsilon_T\leq\varepsilon[m]~\&~\gamma_{SR}<\gamma_0$},\\
\lambda_{\rm{III}}, &\mbox{if ~$\varepsilon_T\leq\varepsilon[m]~\&~\gamma_{SR}\geq\gamma_0$},
\end{cases}
\end{equation}
where $\varepsilon[m]$ denotes the relay's residual energy at the beginning of the $m$-th transmission block. Moreover, the residual energy at the beginning of the $(m+1)$th transmission block can thus be expressed as
\begin{equation}
\varepsilon[m+1] =
\begin{cases}
\min \{\varepsilon[m] + \varepsilon_H^{\rm{I}}, C \}, &\mbox{if $\lambda[m] = \lambda_{\rm{I}}$}\\
\min \{\varepsilon[m] + \varepsilon_H^{\rm{II}}, C \}, &\mbox{if $\lambda[m] = \lambda_{\rm{II}}$}\\
\varepsilon[m] - \varepsilon_T, &\mbox{if $\lambda[m] = \lambda_{\rm{III}}$}\\
\end{cases}.
\end{equation}

Based on the above mathematical description, we now derive the state transition probabilities of the formulated MC for the relay's battery. Inspired by \cite{krikidis2012rf}, the state transition of the MC can be generally split into the following eight cases.

\subsubsection{The battery remains empty ($S_0$ to $S_0$)}
We consider the MC starts with the state $S_0$, i.e., the battery of $R$ is empty. It is obvious that Mode I will be activated in this case. Furthermore, the amount of harvested energy during the current block should be discretized to zero, which indicates that the condition $E_H^{\rm{I}} < \varepsilon_1=C/L$ holds. The transition probability of this case is characterized as
\begin{equation}
\small{
P_{0,0} = \Pr\left\{E_{H}^{\rm{I}} < \frac{C}{L} \right\}=F_{H_{SR}}\left(\frac{C}{\eta P_S L}\right),
}\end{equation}
where $F_{H_{SR}}(\cdot)$ denotes the cumulative distribution function (CDF) of $H_{SR}$. According to \cite{ko2000average}, we can write the CDF of $H_{SR}$ as $F_{H_{SR}}\left(x\right) = 1-Q_N\left(\sqrt{2NK},\sqrt{\frac{2(K+1)}{\Omega_{SR}}x}\right)$,
where $Q_N\left(\cdot,\cdot\right)$ is the generalized ($N$th-order) Marcum $Q$-function \cite{zwillinger2014table}, $K$ is the Rician $K$-factor defined as the ratio of the powers of the LoS component to the scattered components and $\Omega_{SR} = \mathbb{E}\{\left|h_{SR,i}\right|^2\},~\forall i\in\left\{1,\ldots,N\right\}$, with $\mathbb{E}\{\cdot\} $ denoting the statistical expectation and $h_{SR,i}$ denoting the $i$-th element of ${\bf h}_{SR}$.
\subsubsection{The empty battery is partially charged ($S_0$ to $S_i$ with $0<i<L$)}
Mode I is activated to charge the battery. We can also deduce that the effective amount of harvested energy should be expressed as $\varepsilon_H^{\rm{I}} = i C/L$, which means $E_H^{\rm{I}}$ falls between the battery levels $i$ and $i+1$. Thus, the transition probability is
\begin{equation}
\small{
\begin{split}
P_{0,i} &= \Pr\left\{\frac{i C}{L} \leq E_{H}^{\rm{I}} < \frac{\left(i+1\right)C}{L}\right\} \\
        &= F_{H_{SR}}\left(\frac{\left(i+1\right)C}{\eta P_S L}\right) -  F_{H_{SR}}\left(\frac{i C}{\eta P_S L}\right).
\end{split}}
\end{equation}

\subsubsection{The empty battery is fully charged ($S_0$ to $S_L$)}
Similar to the previous two cases, the transition probability can be calculated as
\begin{equation}
\small{
P_{0,L} = \Pr\left\{E_{H}^{\rm{I}} \geq C \right\} = 1 - F_{H_{SR}}\left(\frac{C}{\eta P_S}\right).
}\end{equation}

\subsubsection{The non-empty and non-full battery remains unchanged ($S_i$ to $S_i$ with $0<i<L$)}\label{ii_section}
The battery stays at the same level, which indicates that $R$ either operates in Mode I or Mode II with zero effective harvested energy (i.e. $E_{H}^{\rm{I}}$ or $E_{H}^{\rm{II}}$ is discretized to zero). The transition probability of this case is characterized as
\begin{equation}\label{ii}
\small{
\begin{split}
&P_{i,i} = \Pr\left\{ \left[ \left(E_T > \frac{i C}{L}\right) \cap \left( E_{H}^{\rm{I}} < \frac{C}{L}\right) \right]\right.\\
 &~~~~~\left.\cup \left[\left(E_T \leq \frac{i C}{L}\right) \cap  \left(\gamma_{SR} < \gamma_0\right) \cap  \left( E_{H}^{\rm{II}}< \frac{C}{L}\right)\right] \right\}  \\
        &=
        \begin{cases}
        F_{H_{SR}}\left(\frac{C}{\eta P_S L}\right),~~   \mbox{if ~$E_T > \frac{i C}{L}$};\\
        F_{H_{SR}}\left(\frac{\gamma_0 N_0}{P_S}\right) ,~~\mbox{if ~$E_T \leq \frac{i C}{L}~\&~\gamma_0 N_0 < \frac{2 C}{\eta L}$};   \\
        F_{H_{SR}}\left(\frac{2 C}{\eta P_S L}\right), ~~\mbox{if ~$E_T \leq \frac{i C}{L}~\&~\gamma_0 N_0 \geq \frac{2 C}{\eta L}$} .   \\
        \end{cases}
\end{split}}
\end{equation}

\subsubsection{The non-empty battery is partially charged ($S_i$ to $S_j$ with $0<i<j<L$)}
Similar as the previous case, the battery is partially charged from level $i$ to $j$ (i.e., $\varepsilon_H^X = \left(j-i\right)C/L$). Thus, the transition probability can be derived as (\ref{ij}) shown on the top of next page.
\begin{figure*}[!t]
\vspace*{4pt}
\normalsize
\setcounter{MYtempeqncnt}{\value{equation}}
\begin{equation}\label{ij}
\small{
\begin{split}
P_{i,j} = & \Pr\left\{ \left[ \left(E_T > \frac{i C}{L}\right) \cap \left( \frac{\left(j-i\right)C}{L} \leq E_{H}^{\rm{I}} < \frac{\left(j-i+1\right)C}{L}\right) \right]\right.\\
               &~~~~~\left.\cup\left[\left(E_T \leq \frac{i C}{L}\right) \cap \left(\gamma_{SR} < \gamma_0\right) \cap  \left( \frac{\left(j-i\right)C}{L}\leq E_{H}^{\rm{II}} < \frac{\left(j-i+1\right)C}{L}\right)\right] \right\}\\
        = &
        \begin{cases}
        F_{H_{SR}}\left(\frac{\left(j-i+1\right)C}{\eta P_S L}\right)-F_{H_{SR}}\left(\frac{\left(j-i\right)C}{\eta P_S L}\right),
            &\mbox{if ~$E_T > \frac{i C}{L}$;}            \\
        0,   &\mbox{if ~$E_T \leq \frac{i C}{L}~\&~ \gamma_0 N_0 < \frac{2 \left(j-i\right)C}{\eta L}$;}            \\
        F_{H_{SR}}\left(\frac{\gamma_0 N_0}{P_s}\right)-F_{H_{SR}}\left(\frac{2\left(j-i\right)C}{\eta P_S L}\right),
            &\mbox{if ~$E_T \leq \frac{i C}{L}~\&~ \frac{2 \left(j-i\right)C}{\eta L} \leq \gamma_0 N_0 < \frac{2 \left(j-i+1\right)C}{\eta L}$;}             \\
        F_{H_{SR}}\left(\frac{2\left(j-i+1\right)C}{\eta P_S L}\right)-F_{H_{SR}}\left(\frac{2\left(j-i\right)C}{\eta P_S L}\right),
            &\mbox{if ~$E_T \leq \frac{i C}{L}~\&~ \gamma_0 N_0 \geq \frac{2 \left(j-i+1\right)C}{\eta L}$.}          \\
        \end{cases}
\end{split}}
\end{equation}
\hrulefill
\vspace*{4pt}
\end{figure*}

\subsubsection{The non-empty and non-full battery is fully charged ($S_i$ to $S_L$ with $0<i<L$)}
In this case, the effective harvested energy $\varepsilon_H^X$, either from Mode I or Mode II, is greater than the residual space of the battery. The transition probability is thus given by
\begin{equation}\label{iL}
\small{
\begin{split}
&P_{i,L} = \Pr\left\{ \left[ \left(E_T > \frac{i C}{L}\right)\cap \left( E_{H}^{\rm{I}} \geq \frac{\left(L-i\right)C}{L}\right) \right]\right.\\
               &\left.\cup\left[\left(E_T \leq \frac{i C}{L}\right) \cap  \left(\gamma_{SR} < \gamma_0\right) \cap  \left( E_{H}^{\rm{II}}\geq \frac{\left(L-i\right)C}{L}\right)\right] \right\}\\
        &=
        \begin{cases}
        1-F_{H_{SR}}\left(\frac{\left(L-i\right)C}{\eta P_S L}\right),
        ~\mbox{if $E_T > \frac{i C}{L}$;}    \\
        0,~~~~~~~~~~~~~~~~~~~~~~~~\mbox{if $E_T \leq \frac{i C}{L}~\&~\gamma_0 < \frac{2\left(L-i\right) C}{\eta N_0 L}$;}~~~~~~~~~~~~~~~~~~~    \\
        F_{H_{SR}}\left(\frac{\gamma_0 N_0}{P_S}\right)-F_{H_{SR}}\left(\frac{\left(L-i\right)C}{\eta P_S L}\right),\\
        ~~~~~~~~~~~~~~~~~~~~~~~~~~~\mbox{if $E_T \leq \frac{i C}{L}~\&~\gamma_0 \geq \frac{2\left(L-i\right) C}{\eta N_0 L}$.}\\
        \end{cases}
\end{split}
}\end{equation}

\subsubsection{The battery remains full ($S_L$ to $S_L$)}
In this case, the battery of $R$ certainly has enough energy to support information forwarding in the second hop. Thus, only Mode II can be performed so that the battery level is not reduced. Since the battery is full at the beginning of the transition, $\varepsilon_H^{\rm{II}}$ can be any value. The transition probability can be evaluated~as
\begin{equation}\label{LL}
\small{
\begin{split}
P_{L,L} =  \Pr\left\{\gamma_{SR} < \gamma_0 \right\}
        =  F_{H_{SR}}\left(\frac{\gamma_0 N_0}{P_S}\right)
\end{split}}.
\end{equation}

\subsubsection{The non-empty battery discharged ($S_j$ to $S_i$ with $0\leq i< j\leq L$)}
According to the principle of the proposed ATF scheme described in Sec. \ref{System}, the battery level is decreased only when Mode III is operated. The transition probability can thus be evaluated as
\begin{equation}\label{ji}
\small{
\begin{split}
P_{j,i} =& \Pr\left\{\left(\gamma_{SR} > \gamma_0\right) \cap \left(E_T = \frac{\left(j-i\right)C}{L}\right)\right\}       \\
        =&
        \begin{cases}
        1-F_{H_{SR}}\left(\frac{\gamma_0 N_0}{P_S}\right),
        &\mbox{if ~$E_T = \frac{\left(j-i\right)C}{L}$;}        \\
        0,
        &\mbox{if ~$E_T \neq \frac{\left(j-i\right)C}{L}$.}        \\
        \end{cases}
\end{split}}
\end{equation}

We are now ready to derive the steady state distribution of the relay battery. Let $\mathbf{M} = \left[P_{i,j}\right]_{(L+1)\times(L+1)}$ denote the state transition matrix of the formulated MC. It is easy to verify that $\mathbf{M}$ is irreducible and row stochastic. Thus, there should exists a unique solution $\boldsymbol{\pi}$ that satisfies the following equation~\cite{krikidis2012buffer}
\begin{equation}
\mathbf{\boldsymbol{\pi}} = \left(\pi_0,\pi_1,\ldots,\pi_L\right)^T=\mathbf{M}^T \boldsymbol{\pi}.
\end{equation}
This $\boldsymbol{\pi}$ is actually the discrete distribution of the relay residual energy and can be calculated as
\begin{equation}
\boldsymbol{\pi} = \left(\mathbf{M}^T - \mathbf{I} + \mathbf{B}\right)^{-1}\mathbf{b},
\end{equation}
where $\mathbf{M}^T$ denotes the transpose matrix of $\mathbf{M}$, $\mathbf{I}$ is the identity matrix, $B_{i,j} = 1, \forall{i,j}$, and $\mathbf{b} = \left(1,1,\ldots,1\right)^T$\cite{krikidis2012buffer}.

\subsection{Outage Probability}
Based on the steady state of the relay battery derived in the previous subsection, we now analyze the outage probability of the proposed ATF scheme. Let $\Phi_X$, $X\in\left({\rm{I}}, {\rm{II}}, {\rm{III}}\right)$ denote the outage event of Mode I, II, and III, respectively. According to the full probability theory, we can express the outage probability of the considered WPCCN as
\begin{equation}\label{Outage}
\small{
\begin{split}
P_{\rm{out}}&=\left(1-P_E\right) \Pr\left\{\Phi_{\rm{I}}\right\} + P_E \Pr\left\{\gamma_{SR} < \gamma_0 \right\}\Pr\left\{\Phi_{\rm{II}}\right\} \\
&~~~+ P_E \Pr\left\{\gamma_{SR}\geq\gamma_0\right\} \Pr\left\{\Phi_{\rm{III}}\right\},
\end{split}
}\end{equation}
where $P_E$ denotes the probability that the residual energy at $R$ is no less than the energy threshold $E_T$, which can be expressed as
\begin{equation}\label{P_E}
P_E = \sum\nolimits_{i=k}^{L} \pi_i,~{\rm{s.t.}}~k = \arg\min\nolimits_{k\in1,\ldots,L}\left\{\varepsilon_k \geq E_T\right\}.
\end{equation}

In Mode I, $S$ sends the information to $D$ during the whole block without the help of $R$. Thus, we have
\begin{equation}\label{phi1}
\Pr\left\{\Phi_{\rm{I}}\right\} = \Pr\left\{\gamma_{D}^{\rm{I}} < \gamma_1\right\} = F_{H_{SD}}\left(\frac{\gamma_1 N_0}{P_S}\right),
\end{equation}
where $\gamma_1 = 2^\mathbb{R}-1$ is the outage threshold without cooperation and $F_{H_{SD}}(\cdot)$ is the CDF of $H_{SD}$. Since the $S$-$D$ link suffers from Rayleigh fading, we have
$F_{H_{SD}}\left(y\right) = 1-  \exp \left({-\frac{y}{\Omega_{SD}}}\right)$,
where $\Omega_{SD} = \mathbb{E}\{\left|h_{SD}\right|^2\}$ is the mean of $H_{SD}$\cite{louie2008performance}.

For Mode II, the outage probability can be characterized as
\begin{equation}\label{phi2}
\Pr\left\{\Phi_{\rm{II}}\right\} =\Pr\left\{\gamma_{D}^{\rm{II}}<\gamma_0\right\}= F_{H_{SD}}\left(\frac{\gamma_0 N_0}{2 P_S}\right).
\end{equation}

Similarly, for Mode III, we can evaluate the outage probability as follows
\begin{equation}\label{phi3}
\small{
\Pr\left\{\Phi_{\rm{III}}\right\} = \Pr\left\{\gamma_{D}^{\rm{III}} < \gamma_0 \right\}= \Pr\left\{\gamma_{RD} + \gamma_{SD} < \gamma_0 \right\}.
}\end{equation}
With the aid of \cite{tannious2008spectrally}, we can express the term $\Pr\left\{\gamma_{RD} + \gamma_{SD} < \gamma_0 \right\}$ in closed-form as (\ref{sum_SNR_TAS_CDF}) on top of next page, in which $\bar{\gamma}_{SD}=P_S \Omega_{SD}/N_0$ and $\bar{\gamma}_{RD}=2E_T \Omega_{RD}/N_0$
and $\Omega_{RD} = \mathbb{E}\{\left|h_{RD,i}\right|^2\},~\forall i\in\left\{1,\ldots,N\right\}$ with $h_{RD,i}$ denoting the $i$-th element of ${\bf h}_{RD}$.
By substituting (\ref{P_E}), (\ref{phi1}), (\ref{phi2}) and (\ref{phi3}) into (\ref{Outage}), we now have obtained a closed-form expression for the outage probability of the proposed ATF~protocol.
\begin{figure*}[!t]
\vspace*{4pt}
\normalsize
\setcounter{MYtempeqncnt}{\value{equation}}
\begin{equation}\label{sum_SNR_TAS_CDF}
\Pr\left\{\gamma_{RD} + \gamma_{SD} < \gamma_0 \right\} = N \sum_{k=0}^{N-1}\left( {\begin{array}{*{20}{c}}
N\\
k
\end{array}} \right)\frac{\left(-1\right)^k \left[\bar{\gamma}_{SD}\left(1-\exp{\left(-\frac{\gamma}{\bar{\gamma}_{SD}}\right)}\right)
-\frac{\bar{\gamma}_{RD}}{k+1}\left(1-\exp{\left(-\frac{\left(k+1\right)\gamma}{\bar{\gamma}_{RD}}\right)}\right)\right]}{\left(k+1\right)\bar{\gamma}_{SD}-\bar{\gamma}_{RD}}.
\end{equation}
\hrulefill
\vspace*{4pt}
\end{figure*}
\section{Numerical Results}
In this section, we provide some simulation results to verify the above theoretical analysis and illustrate the impacts of several parameters on system performance. We adopt the channel model $\Omega_{ij}=\left(1+d_{ij}^{\alpha}\right)^{-1}$ to capture the path-loss effect, where $d_{ij}$ denotes the distance between nodes $i$ and $j$, $\alpha\in\left[2,5\right]$ is the path-loss exponent. In the following simulations, we set $d_{SD}=50$m, $d_{SR}=5$m, $d_{RD}=45$m, the path-loss factor for all paths $\alpha=3$, the Rician-factor $K=10$, the noise power $N_0=-60$ dBm, the energy conversion efficiency $\eta=0.5$, and the transmission rate of the system $\mathbb{R}=1$.

\begin{figure}[!t]
  \centering\scalebox{0.5}{\includegraphics{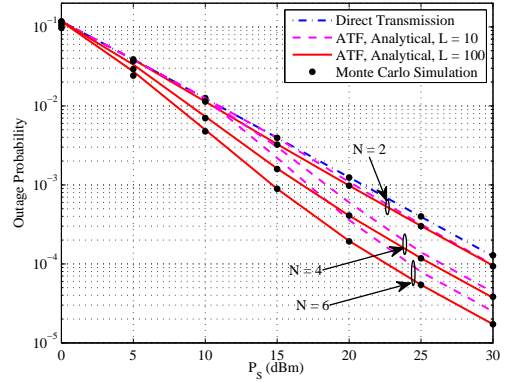}}
  \caption{Outage probability of the proposed ATF scheme versus the source transmit power for different antenna numbers and battery levels, where $\mathbb{R}=1$, $C=5\times10^{-3}$, $E_T=1\times10^{-4}$, $N=[2, 4, 6]$.}\label{Final1}
\end{figure}
\begin{figure}[!t]
  \centering\scalebox{0.5}{\includegraphics{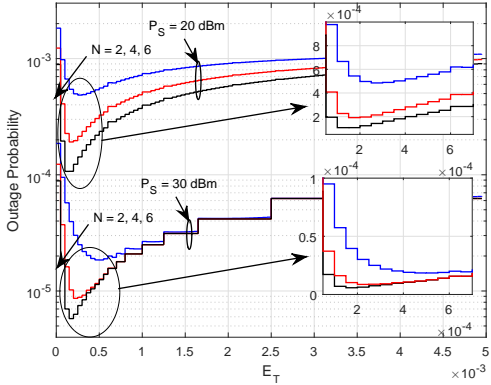}}
  \caption{Outage probability of the proposed ATF scheme versus the consumed energy for information forwarding at the relay with different transmit power~at the source, where $\mathbb{R}=1$, $C=5\times10^{-3}$,~$P_S = [20,30]$dBm,~$N=[2,4,6]$}\label{Final2}
\end{figure}

We first compare the analytical outage probability of the considered system with its associated Monte Carlo simulation, which corresponds to the practical case that the
charging of the relay batteries is continuous (i.e., $\ L \to \infty$). To this end, in Fig. \ref{Final1}, we plots the outage probability curves of the proposed ATF scheme versus the source transmit power for different antenna numbers and battery levels. It can be seen from this figure that the derived analytical expressions of outage probability approach its corresponding
Monte Carlo simulation results as the discrete battery level $L$ increases. Particularly, when
$L = 100$, the analytical expression coincides well with its corresponding simulation, which verifies the effectiveness of the adopted MC model and the correctness of our
theoretical analysis in Sec. \ref{Analysis}.
We can also observe from Fig. \ref{Final1} that when the source transmit power $P_S$ is small, the outage probability of the proposed ATF scheme is similar to that of the direct transmission without cooperation. This is because the relay cannot accumulate sufficient energy to forward information and $R$ keeps operating in Mode I. However, as $P_S$ increases, the proposed ATF gradually outperforms the direct transmission scheme as the relay can accumulate enough energy such that it can assist the source's transmission opportunistically.
Furthermore, the more the antennas of the relay, the larger the performance gap introduced by larger battery levels.

The outage probability of the ATF scheme versus the transmit energy at the relay (i.e, $E_T$) is drawn for different source transmit powers and antenna numbers in Fig. \ref{Final2}. This figure is a stair-stepping plot due to the adopted discrete battery model. It can be observed from the figure that there exists an optimal value of $E_T$, which minimizes the system outage probability. Moreover, when the source transmit power increases from 20dBm to 30dBm, the optimal transmit energy of the relay slightly shifts to the right as the relay can harvest more energy. Furthermore, for a fixed source transmit power, the relay with more antennas requires a smaller optimal transmit~energy.
\begin{figure}[!t]
  \centering\scalebox{0.5}{\includegraphics{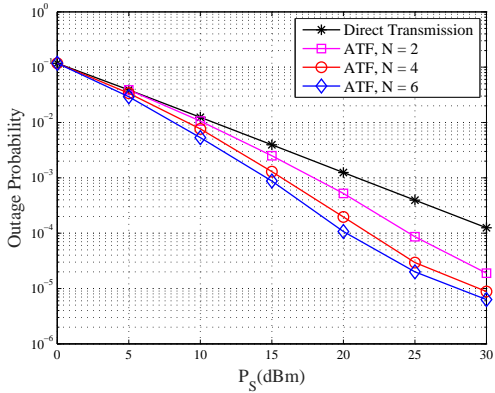}}
  \caption{Outage probability of the proposed ATF scheme with optimal transmit energy at the relay, where $\mathbb{R}=1$, $C=5\times10^{-3}$, $N=[2, 4, 6]$.}\label{Final3}
\end{figure}

Fig. \ref{Final3} performs the outage performance comparison between the proposed ATF protocol with optimal transmit energy at the relay and the direct transmission scheme without cooperation. Note that the optimal transmit energy at the relay for a certain setup could be readily obtained via an exhaustive search of all discrete energy levels. We can see from this figure that with the optimal $E_T$, the proposed ATF protocol is significantly superior to the direct transmission scheme, especially when the source transmit power is high enough. In addition, this performance gain can be further enlarged by increasing the number of antennas equipped at the relay. This is because equipping more antennas at relay can not only effectively increase the amount of harvested energy in the first hop but also efficiently improve the received SNR at the destination in the second hop.

\section{Conclusion}
In this paper we developed an accumulate-then-forward (ATF) protocol for cooperative communications via a multi-antenna energy harvesting relay. By modeling the charging/discharging behaviors of the relay battery as a finite-state Markov chain, we derived a closed-form expression for the exact outage probability of the considered network over mixed Rician-Rayleigh fading channels. Numerical results showed that the system outage probability decreases with the increase of source transmit power and number of antennas at relay. Furthermore, the proposed ATF protocol can outperform the direct transmission scheme, especially when the relay consumes the optimal amount of energy for information forwarding.

\bibliographystyle{IEEEtran}
\bibliography{energy_harvesting_endnote}

\end{document}